\documentclass[pra,twocolumn,floatfix]{revtex4}

\newcommand{\be}{\begin{eqnarray}}
\newcommand{\ee}{\end{eqnarray}}

\newcommand{\vB}{{\bf B}}
\newcommand{\vE}{{\bf E}}
\newcommand{\vvr}{{\bf r}}
\newcommand{\va}{{\bf a}}
\newcommand{\vf}{{\bf f}}
\newcommand{\vk}{{\bf k}}
\newcommand{\vn}{{\bf n}}
\newcommand{\vu}{{\bf u}}
\newcommand{\vv}{{\bf v}}
\newcommand{\vx}{{\bf x}}

\newcommand{\dby}[2]{\frac{{\rm d} #1}{{\rm d} #2}}

\newcommand{\myfig}[2]
{\centerline{\resizebox{!}{#1\textwidth}{\includegraphics{#2}}}}

\usepackage{graphicx}
\usepackage{amssymb}

\begin{document}

\title{Tracking the radiation reaction energy when charged bodies accelerate}

\author{Andrew M. Steane}
\email{a.steane@physics.ox.ac.uk} 
\affiliation{Department of Atomic and Laser Physics, Clarendon Laboratory, Parks Road, Oxford OX1 3PU, England.}


\begin{abstract}
We address some questions related to radiation and energy conservation in classical electromagnetism.
We first treat the well-known problem of energy accounting during radiation from a uniformly accelerating particle.
We present the problem in the form of a paradox, and then answer it using a modern treatment
of radiation reaction and self-force, as it appears in the expression due to Eliezer and Ford and O'Connell. We
clarify the influence of the Schott force and the
total radiated power, which differs from Larmor's formula. Finally,
we present a simple and highly visual argument which enables one to
track the radiated energy without the need to appeal to the far field in the distant future
(the `wave zone').
\end{abstract}
\maketitle


This paper discusses the physics of an accelerating electric charge, with particular regard to
the emitted radiation and the radiation reaction force. In the first part we present 
some issues concerning radiation reaction that often confuse students. As an appealing
way in to the subject, we start with a paradox that has been treated before but perhaps
is less well known than it might usefully be. A well-framed paradox, and its resolution,
offers a helpful way to capture a physical idea, and is often easily memorable for students.
In the second part of the paper we invoke some very simple and easily visualized arguments
to get further understanding of the radiated power.

The paradox concerns the radiation of a charge undergoing uniform acceleration. When
students first learn about radiation reaction, this situation commonly causes confusion,
because the self-force vanishes, which appears to imply either that there is no radiated
energy, or, if there is, then energy conservation has broken down (as we elaborate below).
Indeed this situation caused confusion in the professional physics community for a very
long time, even though the essential insight was already given by Schott in 1915 \cite{15Schott}.
Since 1960, and perhaps earlier, there has not been good reason to
doubt that, when observed by an inertial observer, a uniformly accelerating charge
radiates \cite{98Schwinger,60Fulton,00EriksenII,90Rohrlich}, but the right way to 
describe self-force and radiation reaction has remained unclear until at least
1991 \cite{91Ford,14Burton,14SteaneB,14SteaneA}. The work done by the self-force was 
described by Ford and O'Connell \cite{91FordII,11Intravaia}, who thus obtained the
formula for the radiated power, which is slightly different to Larmor's formula
owing to the fact that the radiating charge cannot be truly point-like.
Here we describe both this and the work done by the Schott term 
either differently or somewhat more
fully than was done previously \cite{02Rohrlich,06Heras,07Hnizdo}. Our treatment is informed by
a discussion of Rohrlich who did the equivalent analysis based on
a slightly different formula for the self-force\cite{00RohrlichA}.
We find the radiated energy per unit time taken to
emit it is given by
\be
\frac{2 q^3}{3 c^3} \frac{ \dot{v}_\lambda f^\lambda } {m}  \label{radpower}
\ee
(Eq. (\ref{P1}) below). This differs slightly from the formula given by Ford and O'Connell,
but we find that difference is insignificant, as we discuss in section \ref{s.react},
whereas the difference from Larmor's formula (Eq. (\ref{PLarmor})) is significant. Equation
(\ref{radpower}) was previously noted in \cite{02Rohrlich} and, at low velocities,
in \cite{07Hnizdo}.

In the second half of the paper we
introduce a convenient method to understand the radiant energy in the field, which avoids
the need to consider the details of the field either close to or far from the particle.
We obtain the radiated power from the field energy, and show that it agrees with
the conclusions obtained from the self-force.

The two parts of the paper have in common that they concern energy movements in the
electromagnetic field, and they offer ways to calculate the power in the electromagnetic 
radiation without the need
to invoke the `wave zone'. For any given event at which a charge is accelerating, the wave
zone is the exterior of
a large spherical surface centred on the event's position in a given frame, at a 
time in the distant future such that it lies on the future
light cone of the event. This surface is usually invoked (either explicitly or implicitly)
in derivations of Larmor's
formula for the emitted power \cite{98Jackson,12Steane}, but it is useful to note that the formula
can be obtained without appealing to that abstraction.

\section{An energy paradox}

We will consider a charged object whose spatial size is small compared to most other relevant
distances in the problem, so we will refer to it as a `particle', though it should not be assumed
that this `particle' is truly point-like, only small \cite{91Ford,61Erber,14SteaneB}. The precise shape
of the object is not important, but if we take it to be roughly spherical then its radius $R$ must
exceed $\sim \! q^2/m c^2$, where $q$ is its charge and $m$ its observed mass,
and we assume the motion under discussion has an acceleration small compared to
$c^2/R$. 

Consider such a particle moving at constant velocity $v$ in the positive
$x$-direction, and also consider a similar particle moving initially at the speed $v$ in the opposite
direction, and then undergoing motion at constant proper acceleration in the positive $x$ direction, 
until it attains velocity $v$ in the positive $x$-direction, after which it moves
at constant velocity (figure 1).

\begin{figure}
\myfig{0.2}{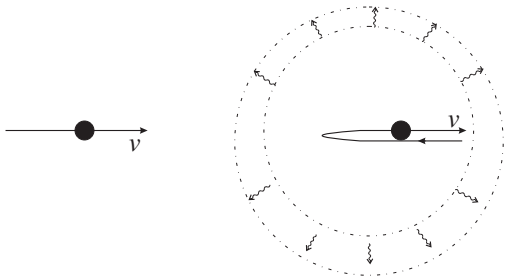}
\caption{A pair of small charged objects with the same final state of motion; one has accelerated, one has not.}
\label{fig1}
\end{figure}

In these two scenarios, the initial and final kinetic energies of the particles are the same.
Also, the initial electromagnetic field is the same in the two cases, up to a translation and
a sign change in the magnetic part.
The final electromagnetic field is not the same, because in the second case there is 
electromagnetic radiation, in the first case there is not. In the second case, 
as the radiation propagates outwards, the field becomes identical to that in the first case 
throughout a larger and larger region of space, and the radiated pulse conserves
its own energy as it propagates. Therefore, the total energy in the final electromagnetic
field in the second case is greater than that in the first, by the energy $W$ in the radiated
pulse. It also follows that the net change in field energy, between initial and final conditions
in the second case, is equal to $W$. 

The question is, where has this energy $W$ come from?

Consider the work done by the applied force $\vf$. (To be clear, throughout this paper, the symbol
$\vf$ (and $f$) without label refers to the external force which is applied to the particle in question
and is not caused by any field sourced by the particle).
The exact relativistic equation of motion is
\be
\vf  + \vf_{\rm self} = \dby{}{t}\left( \gamma m \vv \right)         \label{eqmotion}
\ee
where $\vf_{\rm self}$ is the self-force which is associated with momentum movements in the
field sourced by the charge, such as radiation reaction, and $m$ is the observed rest mass
(which includes a contribution from electromagnetic energy and binding energy). 
Motion at constant proper acceleration has the special property
that the self-force vanishes, $\vf_{\rm self} = 0$ (see Eq. (\ref{f3self}) and section \ref{s.react}).
But if $\vf_{\rm self}$ vanishes in Eq. (\ref{eqmotion}) then the equation is the same as
that describing motion of an uncharged particle, and, in particular,
the total work done by the external force, in the motion under
consideration, is precisely zero (since there is no overall change in the particle's kinetic
energy). In other words
the work done by $\vf$ is just sufficient to give the observed kinetic energy
change of the particle---i.e. zero in total---and no more. Therefore, it would appear,
the external force has not supplied
the radiated energy $W$. So where has the radiated energy come from?

One can see that the radiated energy has not come from the bound field of the
charge in question, because the
final bound field becomes eventually identical to the initial bound field, apart from
a translation and a sign change in the magnetic part. Once again, then, which physical
system has supplied the energy that ends up in the radiation?

In the scenario under consideration, the energy is distributed over an extended
system (the electromagnetic field), whereas energy conservation is enforced locally, 
so perhaps the problem is that we 
have added up the contributions in the wrong way, or over the wrong hyper-surface in
spacetime? Or could it be something to do with the finite spatial extent of the object
and a failure to construct its momentum in the right way? 

Readers who are unfamiliar with this paradox are invited to come to their own conclusions
before reading on.

\subsection{Resolution of the paradox}

The paradox is closely connected to the long-studied question of whether a uniformly
accelerated charge radiates at all \cite{98Schwinger,60Fulton,00EriksenII}. Relative
to an inertial observer, it certainly does, but subtleties
arise when one considers the observations of a uniformly accelerated observer \cite{80Boulware,05Almeida}.
Here we restrict attention to inertial observers, and then
the resolution is simple. The above presentation of the paradox has omitted to consider
the two brief periods when the motion does not have constant proper acceleration, at the beginning
and end of the period of hyperbolic motion. Even though those periods are brief, it turns out that
they contribute
non-negligibly because during them the external force provides all the energy which is eventually
radiated away, as we now show.

For the sake of simplicity, consider the case of low velocities (the `non-relativistic' limit) which
retains all the important features of the paradox. In this limit, the spatial part of the self-force
is (c.f. Eq. (\ref{f4self})) \cite{91Ford,14SteaneB,14Burton}
\be
\vf_{\rm self} = \tau_q \dot{\vf}       \label{f3self}
\ee
where $\tau_q = 2 q^2/3 m c^3$, 
so the equation of motion is
\be
\vf = m \dot{\vv} -  \tau_q \dot{\vf}.   \label{fext}
\ee
If the initial and final speed is $u$ then the acceleration during the hyperbolic motion
is $a = 2 u/T$ where $T$ is
its duration. Let $\delta t$ be the duration of the brief period when the applied force changes
from zero to $m a$, and assume that it also takes this same time $\delta t$ for the force to change
from $m a$ to zero at the end.
Then, during the first such period we have $\dot{\vf} \simeq m\va/\delta t$ and during the second we have
$\dot{\vf} \simeq -m\va/\delta t$  (we shall make a more precise statement in the next section).
The work done by the external force during each period
is approximately $\vf \cdot \vv \delta t$.
Using Eq. (\ref{fext}), this has a part
$m \dot{\vv} \cdot \vv \delta t$ which goes to changing the kinetic energy of the particle,
and a part
\be
\pm \tau_q \frac{m \va}{\delta t} \cdot \vv \delta t
\ee
which contributes energy to the electromagnetic field around the particle. In this equation,
$\va$ is a constant, but $\vv$ is not, and in fact it has opposite sign in the two contributions, so that
they are both equal to
\be
\tau_q m a u = \tau_q m a^2 \frac{T}{2} 
\ee
where we used that the initial and final speed is $u = a T/2$. In other words, in both the initial and the final
periods of changing applied force, $\dot{\vf}$ is in the opposite direction to $\vv$, so the external
force in Eq. (\ref{fext}) has to do some extra positive work, putting energy {\em into} the field, to the total amount
\be
2 \times \tau_q m a^2 \frac{T}{2} = P_{\rm L} T
\ee
where
\be
P_{\rm L} =  \frac{2 q^2}{3 c^3}  \dot{v}_\lambda \dot{v}^\lambda =
\frac{2 q^2}{3 c^3}  a^2          \label{PLarmor}
\ee
is Larmor's formula for the radiated power. We conclude that the external force does do, in total,
just the required amount of work to supply all the radiated energy, and therefore
there is no energy-conservation paradox here. An exact treatment is given in the next section.

The surprise is that the external force provides all this energy
in two brief periods at the start and end of the accelerated motion. Does this mean the particle
is not radiating in between these periods? Not at all. The particle radiates whenever it
accelerates. The energy accounting during accelerated motion has to consider exchange of energy
between the bound
or co-moving field of the charged particle, and the radiated field. During the motion at constant
acceleration, energy is continuously moving from the former to the latter, 
as was first noted by Schott \cite{15Schott}, and as we show explicitly below.

\begin{figure}
\myfig{0.65}{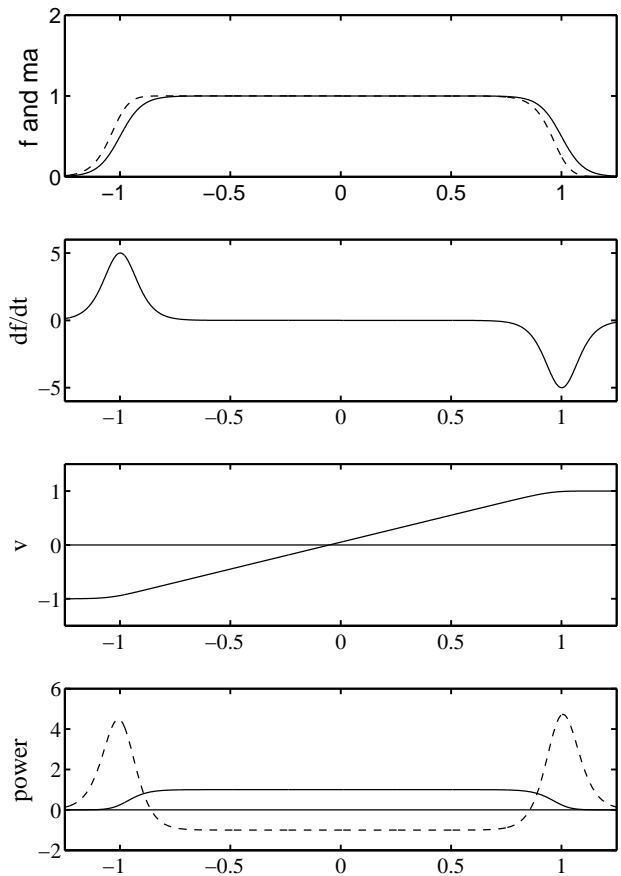}
\caption{A summary of the forces and powers involved in the example motion considered
in the text, each plotted as a function of time. The top graph shows the applied force 
(full line) and resulting $ma$ (dashed line). The next graph shows $df/dt$; the self force
is proportional to this. The third graph shows the particle speed $v(t)$. The last graph shows
two contributions to the power delivered to the electromagnetic field: the radiated power
(full line) and the power delivered to the bound field (dashed line). When $f$ is changing
most of the power goes to the bound field. When $f$ is constant but non-zero the
two contributions are equal and opposite. The total area under the dotted curve is zero.
{\em Remark. The essential insight here is not new, but the author has not been able to
find a presentation like the above in prior work. Pointers to books or papers where the
same idea is set out would be appreciated and will be added to the list of references
as appropriate.}
}
\label{f.eg}
\end{figure}

Figure \ref{f.eg} summarizes the argument by giving the results of an example exact calculation.
We start from a given assumed $\vf(t)$ and then obtain its derivative and hence $\dot{\vv}$
and $\vv(t)$. From this it is easy to extract the work done.

On another matter, note that for this example, and more generally
whenever the force does not change too abruptly,
the acceleration as a function of time (dotted curve in the top graph in figure \ref{f.eg})
is almost the same as the applied force per unit mass
evaluated at a slightly later time (full curve).
This is {\em not} an example of `pre-acceleration'; the equation of motion is strictly causal:
the acceleration at any time is given by the total force $\vf + \tau_q \dot{\vf}$ evaluated 
at that same time, without regard to what may happen at later times.

\subsection{Self-force is not just radiation reaction}        \label{s.react}

The situation of constant applied force, which leads to zero self-force, is easy to misunderstand
because of the common practice of calling the self-force by the name `radiation reaction'.
This is a poor choice of terminology that has misled the physics community for a century.
As Rohrlich rightly
emphasizes \cite{00RohrlichA,90Rohrlich}, it is a mis-nomer because in fact the self-force has three parts. First
there is an `inertial' part describing the supply of 4-momentum to the bound field, which has been
absorbed into the definition of the mass of the particle in our discussion. Next there are
the two terms in the following expression for the self-four-force:
\be
f_{\rm self}^\mu = \tau_q  \left(\dot{f}^\mu - (\dot{v}_\lambda f^\lambda) v^\mu  /c^2\right)  \label{f4self}
\ee
where the dot signifies ${\rm d} / {\rm d}\tau$ (i.e. differentiation with
respect to proper time along the worldline). Readers unfamiliar with this form of the
self-force equation (because, perhaps, they learned the Lorentz-Abraham-Dirac approach)
are referred to \cite{14Burton,14SteaneB,93Ford,48Eliezer}. It is the equation first proposed by
Eliezer and obtained by Ford and O'Connell; it is closely related to but slightly
different from the equation proposed by Landau and Lifshitz. 
The first term in (\ref{f4self}) is the Schott term, it
accounts for redistribution of 4-momentum within the bound field.
The second term describes the supply of power and momentum to the radiated field.
It would be logical to reserve the phrase `radiation reaction' for the second term alone, but it is commonly
applied to both. In this paper we will use the unambiguous phrase `self-force' when discussing
both terms together. The radiated field always transports energy away
from the source, but the bound field may act either to accelerate or decelerate the source,
depending on the recent history of the motion. In physical terms, 
to push a charged particle is to push something that is permanently attached
to a `springy' medium.

Equation (\ref{f4self}) comes from a treatment
which is relativistically consistent but not guaranteed to be exact, because the self-force in
general depends on the shape and internal motion of the accelerating body. However, for
a body of given total charge and not exhibiting extreme behaviours such as 
internal resonance, the corrections to the equation are of higher order in 
powers of $\tau_q$, so for small entities such as electrons, (\ref{f4self}) is very accurate.
The non-relativistic form (\ref{f3self}) follows
by substituting ${\rm d}t$ for ${\rm d}\tau$ and neglecting the second term
in comparison with the first. This does {\em not} amount to neglecting the radiation; this
subtle point is discussed in \cite{00RohrlichA}.

To prove that the self-four-force vanishes 
for motion at constant proper acceleration (hyperbolic motion),
an easy method is to set out to find the motion for which the self-four-force vanishes.
When $f_{\rm self}^\mu = 0$, the equation of motion reads
$f^\mu = m \dot{v}^\mu$ (assuming constant rest mass $m$) so the bracket
on the right hand side of Eq. (\ref{f4self}) is $(m \ddot{v}^\mu - m a_0^2 v^\mu)$
where $a_0$ is the proper acceleration, obtained from $a_0^2 = \dot{v}_\lambda \dot{v}^\lambda$
using a metric signature $(-1,1,1,1)$.
But, the condition $\ddot{v}^\mu = a_0^2 v^\mu$ 
implies hyperbolic motion \cite{12Steane}, so we find $f_{\rm self}^\mu=0$ if and
only if the motion is hyperbolic,
and we also see that this arises by virtue of equal and opposite contributions
from two effects. In the case of hyperbolic motion, during
the initial short period during which $f$ increases from zero to some finite value, the 
external force does more work than is needed to supply the energy eventually required
by the bound field. In the subsequent hyperbolic motion, according to Eq. (\ref{fext})
the applied force does less total work than is needed to supply both the radiated energy and the
kinetic energy of the particle; this is because during such motion 
the bound field near the particle also does work on the particle. While the particle
slows, the system providing the force has work done on it by the particle, and
the bound field holds the particle back a little, tending to maintain its kinetic
energy. As the speed passes through zero this process continues, but now
the system providing the force does work on the particle, and the bound field
`helps' by pulling the particle along a little, doing work on it. At this stage an energy 
deficit is building up: the bound field has less energy than it will eventually
require. This deficit is filled by the applied force during the second period when it changes.
In the second such short period, $\dot{\vf}$ is again opposed to $\vv$ so again the external force
does more work than is needed to supply either kinetic energy or radiated energy;
the energy passes to the bound field and stays there.


We now provide a quantitative statement of the above ideas by calculating the rate of 
doing work by the applied force, in both the general (any $v$) and low-velocity
($v \ll c$) cases.
Our discussion of the various contributions matches that of
Rohrlich \cite{00RohrlichA}, except that we use a different, and better, expression for the self-force.
Previously several authors have treated the radiation power and the Schott power
implied by Eq. (\ref{f4self}); our discussion
slightly extends or modifies the prior ones \cite{07Hnizdo,02Rohrlich,91FordII,06Heras}.

Assuming the rest mass is constant, the relativistic equation of motion is
\be
f^\mu +  \tau_q  \left(\dot{f}^\mu - (\dot{v}_\lambda f^\lambda) v^\mu  /c^2\right) = m \dot{v}^\mu.
\ee
The rate of doing work is given by the zeroth component of this four-force:
\be
\dby{W}{\tau} = f^0 c = m \dot{v}^0 -  \tau_q c \dot{f}^0  + \tau_q \dot{v}_\lambda f^\lambda \gamma
\label{work}
\ee
where $\gamma$ is the Lorentz factor.
The three terms on the right hand side are the rate of change of kinetic energy, the Schott power and the radiated power.
The Schott power takes the form of a total derivative, therefore the net work done by the Schott term, between any 
two events where the four-force has no net change, is zero. The radiation term gives, for the
radiated power per unit time taken to emit it,
\be
P_R = \dby{W_R}{t} = \frac{2 q^3}{3 c^3} \frac{ \dot{v}_\lambda f^\lambda } {m}  \label{P1}
\ee
where we used ${\rm d}t/{\rm d}\tau = \gamma$, and we note that the resulting expression is
Lorentz-invariant. Equation (\ref{P1}) is not quite the same as Larmor's expression (\ref{PLarmor}).
This is because Larmor's expression does not take the finite size of the accelerating body
into account. We will elaborate on this point below and in section \ref{s.pic}.

In the low velocity limit, Eq. (\ref{fext}) gives the rate of doing work
\be
\vf \cdot \vv &=& m \dot{\vv} \cdot \vv -  \tau_q \dot{\vf} \cdot \vv. 
\ee
The first term on the right hand side is the rate of change of kinetic energy of the particle.
To clarify the physical interpretation of the second term, use
\be
\dot{\vf}\cdot \vv &=& \dby{}{t}\left( \vf \cdot \vv\right) - \vf \cdot \dot{\vv}
\ee
so we have
\be
\! \vf \cdot \vv &=& \dby{}{t}\!\left(\! \frac{1}{2} m v^2 \!\right) 
- \tau_q \dby{}{t}  \left( \vf \cdot \vv  \right)  + P   \label{fwork}
\ee
where
\be 
P = \frac{2 q^2}{3 c^3} \frac{\dot{\vv}\cdot \vf} {m} . \label{Pme}
\ee
The three contributions to the rate of doing work 
correspond to the three appearing in the more general expression (\ref{work}).
The radiated power agrees exactly with the expression (\ref{P1}) when $P_R$ is evaluated in
the instantaneous rest frame.

Ford and O'Connell \cite{91FordII} also considered this question based on the same starting point
(\ref{fext}), but they arrived
at a different result for the radiated power:
\be
P_{\rm FO} = \frac{2q^2}{3 c^3} \left( \frac{f}{m} \right)^2 ,  \label{Prad}
\ee
and a more detailed subsequent treatment came to the same conclusion \cite{06Heras}.
In order to understand this, express $\dot{\vv}$ in (\ref{Pme}) in terms of the
force, using Eq.. (\ref{fext}):
$
m \dot{\vv}\cdot\vf = \vf \cdot \vf + \tau_q \dot{\vf} \cdot {\vf}.
$
Therefore
\be
P = P_{\rm FO} + \frac{\tau_q^2}{2m} \dby{}{t}(f^2)
\ee
Hence the two expressions only differ when the size of the applied force is changing, and, furthermore,
they predict the same total radiated power between any two events at which the applied force has
the same size. It follows from this that the choice between $P$ and $P_{\rm FO}$ is largely a matter
of convention, concerning how to apportion the energy between the Schott field and the radiation
while $f$ is changing. We are here making one choice, in agreement with two
previous authors \cite{07Hnizdo,02Rohrlich},
while Heras and O'Connell made the other \cite{06Heras}.
Also, even when $f$ is changing, the two expressions only differ at the next
higher order in $\tau_q$, where our original expression (\ref{f4self}) is not guaranteed to be accurate,
so one should not over-interpret this small difference. This was also noted by Rohrlich \cite{02Rohrlich}.

For the case of a constant force, $\dot{\vf}=0$, the Schott power evaluates 
to $-\tau_q \vf \cdot \dot{\vv} = -\tau_q f^2/m$ and then Eq. (\ref{fwork}) gives
\be
\vf \cdot \vv = \dby{}{t}\left( \frac{1}{2} m v^2 \right) + \frac{2q^2}{3 c^3} \left( \frac{f^2}{m^2} - \frac{f^2}{m^2} \right).
\ee
Here we explicitly exhibit both the radiated power and the power leaving the bound field, for
this case. This helps one to see clearly that in the presence of an applied force,
the radiation is happening throughout the motion, not just when the force is changing.

The overall conclusion is that energy conservation is maintained, and the external force does
indeed supply the energy required by both the bound field and the radiated field. The inertial
contribution to the energy of the bound field has been absorbed into the definition of $m$,
and we have exhibited the other part (the Schott term) explicitly.

\section{Finding radiated energy without recourse to the wave zone}  \label{s.pic}

We now turn to the direct calculation of radiated energy, by examining the field around a
particle which has accelerated.

The standard methods of derivation of Larmor's formula (\ref{PLarmor}) for the power radiated by an
accelerating point charge involve justifying an assumption that only the part of
the field associated with acceleration leads to radiation, and that one may legitimately
calculate the energy associated with this part of the field alone, and call it radiated energy.
One way to justify this is to
take the limit $r \rightarrow \infty$ where $r$ is the distance from
the source event to the field event. For a given source event, the field events in such a
calculation are located on an infinitely large spherical
surface in the infinite future---the `wave zone' or `radiation zone'. 
Sometimes the consideration of this limit is problematic.
In the radiation zone the radiated field carries almost all
the energy, when integrated over all directions, but in some directions it vanishes completely
where the bound field does not, and even where it is strong 
it does not dominate in all respects. For example, its divergence is
everywhere equal and opposite to that of the bound field.
In any case, it is interesting to ask whether one can avoid an 
appeal to the radiation zone. It should, after
all, be possible to learn about something happening in the here and now without recourse to
the far distance and the infinite future.

In a classic paper, Teitelboim \cite{70Teitelboim} addressed this issue
among others, and gave much insight into the energy and momentum movements in
the fields sourced by a charged particle undergoing arbitrary motion. Subsequent work
has further elucidated particular cases or has extended the ideas, for example to non-flat
spacetimes. In the present discussion we wish to give an argument which, owing to its
visual nature and great simplicity, might be useful as a teaching aid. The aim of the
argument is to get some general insight into the movement of energy in an electromagnetic
field, and to derive Larmor's formula without invoking the wave zone.

\begin{figure*}
\myfig{0.35}{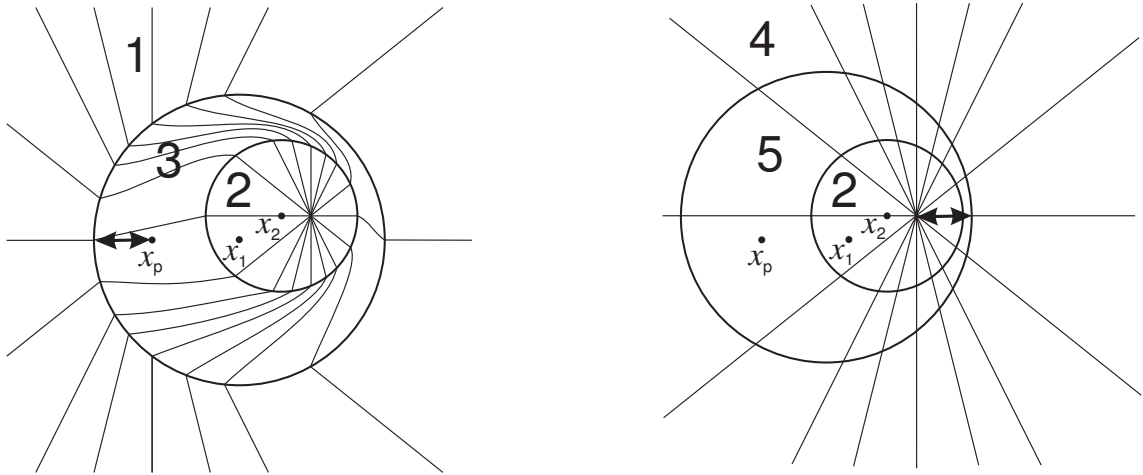}
\caption{The left diagram shows the electric field around a particle which has recently accelerated
and now undergoes uniform motion to the right at a velocity equal and opposite to the one
it had before $t_1$. The right diagram shows the eletric field of
a particle which has the same final state of motion, but has not accelerated. The circles indicate
various regions defined in the text. The bold double-ended arrow has the same length and direction
in the two diagrams. This ensures that the electromagnetic field in region 1 is everywhere 
the same as that in 
region 4, apart from a translation and reflection. It follows that the energy in region 3 exceeds that in region 5
by an amount that is independent of the moment in time chosen to draw the diagram (see text).}
\label{f.regions}
\end{figure*}

Consider a charged particle which moves initially at some constant velocity and finally
at some constant velocity (not necessarily the same) relative to a given inertial
frame. For any such motion, there exists an inertial frame relative to which the initial
and final velocities are equal and opposite. Adopt this frame, oriented so that
the initial and final velocity is $\mp \vu$ in the $x$-direction, and suppose
that the part of the worldline for
which the motion is arbitrary (but always timelike) extends between 
events $(t_1, \vx_1)$ and $(t_2, \vx_2)$.
At any time $t > t_2$, divide all of space into three regions. Region 1 is the exterior
of a sphere of radius $c(t-t_1)$ centred at $\vx_1$. Region 2 is the interior of
a sphere of radius $c(t-t_2)$ centred at $\vx_2$. Since the worldline is timelike,
these regions do not overlap. Define region 3 as the region between them.
These regions are shown, for an example case, in figure \ref{f.regions}a. By reasoning
about this figure, we will make an important observation about the energy movements
in the field.


In region 1 the electromagnetic field is that of a charge uniformly moving at
the initial velocity; in region 2 the field is that of a charge uniformly moving at the
final velocity; in region 3 the field is more complicated, having both radiative and
bound parts. The electric field in region 2 extends radially outwards from the present
position of the particle. The electric field in region 1 extends radially outwards from
the {\em projected position} given by
\be
\vx_{\rm p} = \vx_1 - \vu (t-t_1).
\ee
This is the position the particle would now have  (at time $t$)
had it continued permanently in its initial state of motion. Since $u < c$
the projected position is located inside the sphere enclosed by region 1.

So far we have simply taken a general look at the form of the electromagnetic field. The
only assumption has been that the particle moves initially and finally at constant velocity,
and for convenience we have adopted the reference frame in which those velocities are equal
and opposite.

Let $U_i$ be the energy contained in the electromagnetic field in region $i$. By conservation of
energy, we have, for all times $t > t_2$,
\be
U_1 + U_2 + U_3 = \mbox{constant}.       \label{sumE}
\ee

Next, consider the case of a particle which has never accelerated, but has always moved in the final
state of motion of the particle under consideration. In other words, this `reference' particle has constant velocity 
$\vu$. Its electric field is illustrated in figure \ref{f.regions}B. 
Let $U^{\rm C}_i$ be the energy in the electromagnetic field of the reference particle, in the three regions
we have identified (the C here stands for `Coulomb'; we shall call the field of a uniformly
moving point charge a `moving Coulomb field', where we have in mind an exact treatment including
Lorentz contraction). Since the fields of the actual particle and
the reference particle
are identical in region 2, clearly $U^{\rm C}_2 = U_2$, but we cannot make any such simple
statement about $U_1$ and $U_3$. 

We next identify a further interesting region. This is the crucial part of the argument. 
This further region, region 4, is the exterior of a sphere of radius $c(t-t_1)$ centred at
\be
\vx_t + \vu (t-t_1)
\ee
where $\vx_t = \vx_2 + \vu(t-t_2)$ is the present position of the reference particle. 
This means the centre of the spherical surface defining region 4
is displaced from the present location of the reference particle by the same amount (but
in the opposite direction)
that the centre of the spherical surface defining region 1 is displaced from the projected position
of the actual particle. 
There are two useful implications. First, because $u<c$, region 4 does not overlap region 2. Secondly,
the electromagnetic field of the reference particle in region 4 is the same as the electromagnetic field
of the actual particle in region 1, except for a displacement and a reflection in a plane normal to the $x$ axis. 
An example of this may be seen by examining the pattern of the
field lines in figure \ref{f.regions}. Such
a displacement and reflection does not change the energy content of the field, therefore
\be
U_1 = U^{\rm C}_4.    \label{equal}
\ee

Finally, define region 5 as that between region 2 and region 4. Since the reference particle is in
a perfectly allowed physical condition in which no energy is being supplied, we must have
\be
U^{\rm C}_4 + U^{\rm C}_2 + U^{\rm C}_5 = \mbox{constant}.
\ee
Subtracting this from Eq. (\ref{sumE}) gives
\be
U_1 + U_2 + U_3 - (U^{\rm C}_4 + U^{\rm C}_2 + U^{\rm C}_5) = \mbox{constant}.
\ee
Now, using $U^{\rm C}_2 = U_2$ and Eq. (\ref{equal}), we obtain
\be
U_3 - U^{\rm C}_5 = \mbox{constant}.
\ee
This simple result, easily arrived at as we have shown, tells us something very interesting about the
electromagnetic field sourced by a particle undergoing arbitrary motion. It says that the energy content
of that field differs from what it would need to be to construct the moving Coulomb field of a particle in
the final state of motion, {\em by an amount that does not change with time}. As time goes on, all
the light spheres we have identified grow, and the energy contents of all the regions change.
But regions 3 and 5 have the interesting property we have identified, which is
\be
U_3(t) = U^{\rm C}_5(t) + U_R          \label{UR}
\ee 
where $U_R$ is independent of time. Since as time goes on, the field around the actual particle
becomes more and more like a moving Coulomb field, we can identify $U_R$ as an energy which
has become detached from the particle. It is the radiated energy.

The argument allows us to make the standard observations about the source of electromagnetic
radiation (for inertial observers in the absence of gravity), namely that accelerated motion always results in radiated
energy, non-accelerated motion never does, and the radiated energy moves outwards from
the source at the speed of light. We can also obtain Larmor's formula, as follows.

The fields of a particle in an arbitrary state of motion are, in Gaussian units,
\be
\vE &=& \frac{{q} r^3}{(\vvr \cdot \vvr_0)^3 c^2}\left[
(c^2 - v^2) \vvr_0 + \vvr \wedge (\vvr_0 \wedge \va) \right],     \label{Efield} \\
 \vB &=& {\bf n} \wedge \vE
\ee
where $\vvr$ is the vector from the source event to the field event,
$\vvr_0 = \vvr - \vv r/c$, $\vn = \vvr/r$ and $\vv, \va$ are velocity and acceleration at the source
event.
For a given source event, in the instantaneous rest frame these simplify to
\be
\vE = {q}\left( \frac{\vn}{r^2} + \frac{ \vn \wedge [ \vn \wedge \va ]}{c^2 r} \right),
\;\;\; \vB = \frac{ -{q}}{c^2 r} \vn \wedge \va .
\ee
The energy density in the field is
\be
u = \frac{1}{8\pi} (E^2 + B^2)
= \frac{ {q}^2}{8\pi} \left( \frac{1}{r^4} + \frac{2 a^2 \sin^2 \theta}{c^4 r^2} \right)
\ee
where $\theta$ is the angle between $\vn$ and $\va$
(in SI units one would have $q^2/4\pi \epsilon_0$ instead of $q^2$ in the last version).
Integrating this over a spherical shell of radius $r$
and thickness $\delta r$ we obtain, for the total field energy in such a shell,
\be
\delta U &=& \int_0^{\pi} \int_0^{2\pi} u r^2 \sin \theta {\rm d} \theta {\rm d} \phi \delta r   \nonumber \\
&=& q^2 \left(\frac{1}{2r^2} + \frac{2 a^2}{3 c^4} \right) \delta r.  \label{dU}
\ee
This $\delta U$ is an example of the energy we have called $U_3$ in the argument above.
To get the radiated energy, we subtract from it $U^{\rm C}_5$ which is the energy in the
Coulomb field in the appropriate shell. This is easily obtained; it is equal to the first term
in Eq. (\ref{dU}). Hence we find
\be
U_R =  \frac{2}{3} \frac{q^2 a^2}{c^4} \delta r.         \label{URfinal}
\ee
The time taken to emit this energy is the time taken for a light sphere to grow from radius
$r$ to $r + \delta r$, so we find that the energy radiated, per unit time taken to emit it, is
as given by Larmor's formula, Eq. (\ref{PLarmor}).

\subsection{Extension to objects of finite size}

Larmor's treatment, and the above treatment, gives the answer for a point charge. 
No point-like object can have a finite charge, however, unless the observed mass tends to
infinity, owing to the contribution from the electromagnetic field energy \cite{61Erber,14SteaneB}. 
Therefore Larmor's formula,
and Eq. (\ref{URfinal}), are only valid in the limit $q \rightarrow 0$. In that limit $\tau_q \rightarrow 0$ and then
(\ref{PLarmor}) agrees with (\ref{Pme}). For a charged entity of finite charge and mass, and therefore
non-zero spatial extent, we should expect a departure from Larmor's formula, and 
Eqs (\ref{Pme}) or (\ref{Prad}) give, to first approximation, what that departure is. 
It can be understood as
a small modification in the energy in the radiation field, owing to the difference between the field
of a small extended object and the field of a point-like object. For a small rigid body, moving
non-relativistically, this can be obtained from eqs (3), (10), (26), (29) of \cite{11Intravaia}.
Eq. (10) of \cite{11Intravaia} gives the squared electric field in the radiation zone as
\be
\left|\vE(\vvr,\omega)\right|^2 = q^2 |f_q(\vk)|^2 \omega^4 \left[ \tilde{R}^2\right] \sin^2\theta /c^4 r^2
\label{ER}
\ee
where $f_q$ is the form factor of the rigid charge distribution, for which a suitable expression is
\cite{91Ford,11Intravaia}
\be
|f_q (\vk)|^2 = \frac{1}{1 + \omega^2 \tau_q^2}
\ee
and $\tilde{\bf R}= \alpha(\omega) \tilde{\vf}$ where $\alpha$ is the linear
response function and $\tilde{\vf}$
is the Fourier transform of the applied force. 
It follows that the radiated power is proportional to
\be
\omega^4 |f_q|^2 |\alpha|^2 | \tilde{\vf} |^2 = | \tilde{\vf} |^2 / m^2
\ee
in agreement with Eq. (\ref{Prad}), where we used the expression
$\alpha =  (-1 + i \omega \tau_q)/m\omega^2$ which describes the response of a free particle
according to (\ref{fext}). The above calculation was presented at greater length in \cite{91FordIII}.
This result does not necessarily offer a reason to prefer
(\ref{Prad}) over (\ref{Pme}) because the difference between them appears at a higher
order in powers of $\tau_q$ than is assumed in the approximations leading to (\ref{ER}).

We note that (\ref{Prad}) can also be reproduced to this order of accuracy by 
replacing $\va \rightarrow \va - \tau_q \dot{\va}$ in the Larmor formula, but this
is an observation not a derivation.

The argument of figure \ref{f.regions} and eqs (\ref{sumE}) to (\ref{UR})
remains valid for the case of an extended body with sufficient symmetry, if 
we adapt it as follows. 
We consider the case where the body moves in such a way that its
initial and final motion is inertial, as before, and we adopt the frame in
which the initial and final velocities are equal and opposite. 
We assume that the body has the same proper size and shape in the initial and final
states, and that it has reflection symmetry in a plane perpendicular to
the $x$ axis (the direction of its velocity change), 
and it has undergone no net rotation.
The time $t_1$ should be taken as the time when some part of the body
first starts to accelerate, and $t_2$ is the time when all parts of the body
have ceased to accelerate. Regions 1,2 and 3 are defined as before using spheres
centred on $\vx_1$ and $\vx_2$. Region 4 is identified by using the projected
position of any point on the body, and placing it relative to the
reference body with the same offset from the corresponding point on that body,
after a reflection through a symmetry plane of the body, orthogonal to the initial (or final)
velocity. Since we assumed the body in question is symmetric under reflection
through such a plane, one may as well use a point on the plane of symmetry,
for example the centroid of the body's mass distribution. The argument now
applies as before.

To conclude, this paper has offered contributions of two types: accurate
statements about radiant energy and self-force, and easily visualized or 
remembered ways of thinking about them. The statements correct or clarify
earlier work (by a modest amount). The physical scenarios offer,
we hope, a useful teaching method, whose ideas are captured in the three
figures.

I thank V. Hnizdo for helpful reactions to an early version of the paper.

\bibliography{selfforcerefs}

\end{document}